# Secure Mobile Payment Architecture Enabling Multi-factor Authentication

Hosam Alamleh
*Computer Science*
*University of North Carolina Wilmington*
Wilmington, North Carolina, USA
hosam.amleh@gmail.com

Ali Abdullah S. AlQahtani
*Computer Systems Technology*
*North Carolina A&T State University*
Greensboro, North Carolina, USA
alqahtani.aasa@gmail.com

Baker Al Smadi
*Computer Science*
*Grambling State Univeresity*
Grambling, Louisiana
bakir_smadi@hotmail.com

*Abstract*—The rise of smartphones has led to a significant increase in the usage of mobile payments. Mobile payments allow individuals to access financial resources and make transactions through their mobile devices while on the go. However, the current mobile payment systems were designed to align with traditional payment structures, which limits the full potential of smartphones, including their security features. This has become a major concern in the rapidly growing mobile payment market. To address these security concerns, in this paper we propose new mobile payment architecture. This architecture leverages the advanced capabilities of modern smartphones to verify various aspects of a payment, such as funds, biometrics, location, and others. The proposed system aims to guarantee the legitimacy of transactions and protect against identity theft by verifying multiple elements of a payment. The security of mobile payment systems is crucial, given the rapid growth of the market. Evaluating mobile payment systems based on their authentication, encryption, and fraud detection capabilities is of utmost importance. The proposed architecture provides a secure mobile payment solution that enhances the overall payment experience by taking advantage of the advanced capabilities of modern smartphones. This will not only improve the security of mobile payments but also offer a more user-friendly payment experience for consumers.

## I. INTRODUCTION

Mobile payment is defined as using mobile devices such as smartphones, PDAs, smart watches, or any Near field communication (NFC) enabled-devices to make payments [1]. It was also define as a financial process involving electronic mobile communication devices to initiate, authorize and complete financial transactions [2]. A successful economic transaction is defined as a business activity using an electronic device connected to a mobile network [3]. The third definition is closest to current digital wallet based mobile payment systems. Mobile payment systems are categorized based on proximity and business model. Proximity payments are grouped based on consumer location, e.g. in-store or remote, while the business model is differentiated by consumer relationships or business-consumer relationships. This paper introduces an architecture with focuses on business to consumer mobile payment technologies. Other mobile payment systems include SMS and QR. But the most popular used among digital wallets is NFC [4]. NFC usage is growing due to the increasing popularity of smartphones and their various applications. Unlike SMS payments, NFC payments are made in-person at a store or compatible terminal by simply bringing the mobile device close to the terminal. This technology has garnered significant attention due to its ease of use for data exchange and its potential for integration into various features, as NFC technology allows for limitless possibilities [5].

Mobile payment systems allow customers to make electronic transactions using their smartphones or other mobile devices. These payments can be made in-store, online, or through mobile apps, and can include purchases made with credit or debit cards, as well as digital wallets and other forms of mobile money. Mobile payments have become increasingly popular in recent years, as they offer a convenient and secure alternative to traditional payment methods. They also enable new forms of commerce, such as mobile banking and peer-to-peer transactions. The utilization of mobile payment systems has been on the rise in the past few years. The mobile payment market was valued at USD 43.11 billion in the year 2021 and USD 55.34 billion in the year 2022 it is expected to grow to USD 587.52 billion by the year 2030 growing at a compound annual growth rate of 37.1 % during the forecast period. [6]. Globally, east Asian countries has high adoption of mobile payment Apps. In the US, 43.2 % of the population use mobile payments. With Apple pay, Google pay, Samsung pay, and Starbucks among the top vendors [7].

In general, mobile payment is considered more secure because extra protection added to the mobile app compared to standard credit card payments. To evaluate a payment system, There are several factors to consider including Authentication, Encryption, and Fraud detection. Authentication is essential to ensure that only authorized users can access the payment functionality. A system should use strong and multi-factor authentication methods, such as fingerprint or facial recogni-tion. Alternately, encryption ensures that payment information data such as credit card numbers and account information are stored securely and transmitted over the network securely. Finally, a good payment system must have risk management features, such as fraud detection and prevention, to identify and stop suspicious activity.

Current mobile payment systems are built to be compatible with existing payment networks infrastructure, primarily for physical card payments using chips or magnetic strips. However, these systems lack robust user authentication, relying mainly on possession of cards and, sometimes, pin numbers. Security is crucial for mobile payments. Despite, standards like PCI DSS (Payment Card Industry Data Security Standard) are employed to maintain the CIA triad, security breaches can still occur, putting personal and payment card information at risk. To address this, a new payment architecture that incorporates authentication methods available in smartphones like location and biometrics, and others is introduced in this paper.

## II. Background

The paper [8]. highlights the growing attention that mobile commerce, specifically mPayment, is receiving from both business and academic communities. The authors note that the proliferation of mobile commerce, especially in the business-to-consumer sector, requires secure and easy-to-use payment methods that are widely available and globally accepted. The authors define mPayment as the ability to make payments using mobile devices such as smartphones and personal digital assistants that are equipped with radio frequency (RF) or near field communication (NFC) technology. The paper recognizes that while mPayment is still in its early stages, its acceptance is expected to grow rapidly in the coming years, particularly in Europe and Asia. However, the authors observe that the adoption of mPayment methods in the US has been slow, largely due to a lack of unified standards, security and privacy concerns, and the slow diffusion of mCommerce. The authors aim to provide a clear understanding of the state of mPayment and to explore the factors that will determine its adoption by US consumers. Moreover, , the paper offers a blueprint for a cross-industry and cross-platform mPayment solution that offers consumers fast and convenient payment processes for both online and in-person transactions. This solution is intended to address the challenges currently facing mPayment adoption in the US and to promote its wider acceptance by consumers.

In the recent years, the interest in mobile payment has grown significantly among consumers, leading to the potential for wider adoption in the near future. This PhD dissertation [9] groups four studies that aim to understand the factors that influence the decision to adopt mobile payments and the variations across different environments and technologies. The research draws from the classic theories of new technology adoption, such as the Technology Acceptance Model (TAM), Theory of Planned Behavior (TPB), and the Theory of Reasoned Action (TRA), as well as more recent theories related to mobile services adoption. The study evaluates four models that incorporate various variables related to both the payments and the users. The research was conducted in Spain, Brazil, and Germany through self-administered web surveys with a sample size of 168, 871, 423, and 2,210 individuals respectively. The findings of this thesis reveal that the models have a high predictive power of the intention to use mobile payments, with values ranging from 56 to 71%. The results indicate that all variables play an important role in the adoption process, but attitude towards use and perceived usefulness stand out as the most significant. Personal innovation is relevant as an antecedent of intention and as a moderator of behavioral intention, and consumer interest in mobile services is also proven to be a relevant precedent of mobile payment adoption.The originality and value of this thesis lie in its contribution to the academic field as one of the first studies to empirically test the determinants of consumer acceptance of payments through QR codes, NFC, and SMS, as well as the role of mobile marketing services in the adoption of mobile payment. Additionally, this study provides a consumer-focused perspective and offers recommendations and strategies to enhance the adoption of mobile payment and integrated mobile payment services.

The paper [10] explores the topic of mobile payment in virtual social networks (VSN) and the factors that determine its level of acceptance by consumers. The study employs a modified version of the classical technological acceptance models (TRA and TAM) to analyze the level of acceptance of mobile payment in VSN. The study proposes an integrated theoretical model named MPAM-VSN which takes into account the relative importance of external influences, ease of use, usefulness, attitude, trust, and risk in terms of the acceptance of mobile payment systems in VSN. The study also analyzes the potential moderating effect of users' experience with similar tools on their acceptance of mobile payment systems in VSN. The empirical results of the study showed that the proposed behavioral model MPAM-VSN was well-adjusted, suggesting that users' previous experience with similar tools increases their intention to use mobile payment systems in VSN. The results of the study have interesting implications for the diffusion of mobile payment systems in VSN.

There are several types of mobile payments types such as SMS, QR and NFC [11]–[14]; however, the most popular used among digital wallets is NFC [15]. NFC, or Near Field Communication, is a type of wireless communication that utilizes RFID technology. It allows for two-way communication between devices within close proximity, usually less than 20 cm. This technology enables consumers to easily make payments by simply holding their mobile device close to a merchant's POS terminal, allowing for the exchange of payment information [16]–[18]. A user may need to enter a secure PIN or password to approve the transaction. It is estimated that NFC mobile payments can be 15 - 30 seconds faster than swiping a traditional card and signing the receipt or entering a PIN [19]. NFC usage is growing due to the increasing popularity of smartphones and their various applications. Unlike SMS payments, NFC and QR payments are made in-person at a store or compatible terminal by simply bringing the mobile device close to the terminal. This technology has garnered significant attention due to its ease of use for data exchange and its potential for integration into various features, as NFC technology allows for limitless possibilities [9]. NFC mobile payments offer several advantages over traditional mobile payment method . For consumers, these advantages include reliability, security, ease of use, convenience, ability to use as a digital wallet, wide acceptance, availability on a variety of devices, and additional value-added applications. [20], and economic attractiveness due to open standards with no licensing fees [15].

In mobile payment systems a card is added to a digital wallet to be used for future transactions. When a credit or debit card is added to a digital wallet, as shown in figure 1, the digital wallet App sends the primary account number (PAN) to the card network, which verifies the PAN with the card issuer. Then, it generates a token to the digital wallet App. This token is then sent to the device to be used for future transaction.

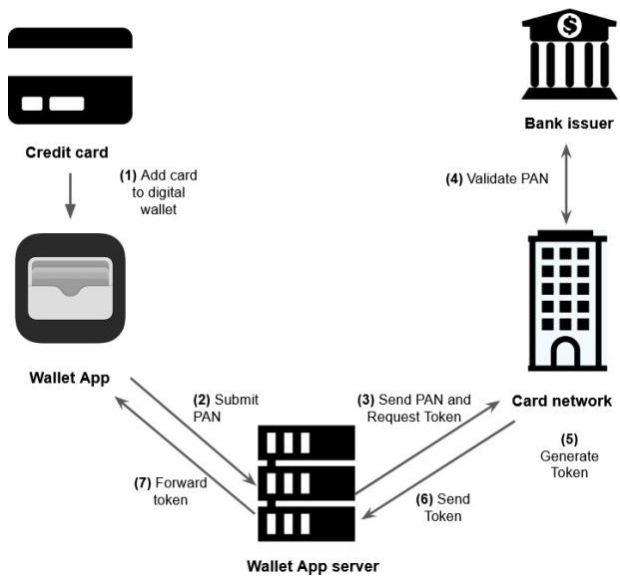

Fig. 1. Mobile payment system-generating a token

When a transaction is initiated, the original token(OT) issued by the card network is used to generate a one-time transaction code (OTTC). As shown in Figure 2. This $OTTC$ is sent from the smartphone to the point of sale (POS) terminal via NFC. The POS then forwards the token to the card network for validation. The card network has a copy of the $OT$ issued to the digital wallet, thus, it will be able to regenerate the $OTTC$ and verify if the received $OTTC$ is authentic . After token validation, the card network verifies the funds availability with the card issuer and sends transaction approval decision the POS terminal. In these systems, $OTTC$ is generated using a (HMAC) function as shown in equation 1:

$$OTTC = HMAC(OT, t) \quad (1)$$

Where $OT$ is the original token issued to the digital wallet by the card network and $t$ is the time of transaction. Including the time in the $OTTC$ generation gives the $OTTC$ a limited validity to defend against replay attacks.

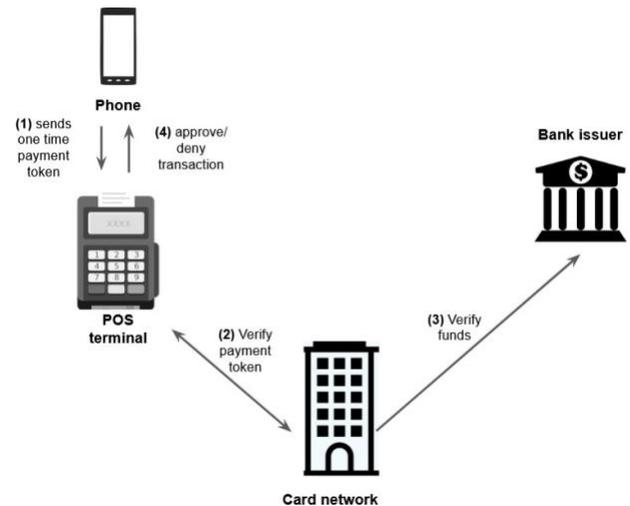

Fig. 2. Transaction overview

As seen from the process discussed above, current mobile payment systems are designed to integrate with existing infrastructure. Current payments systems infrastructure was mainly build to support payments using physical cards that uses chips or magnetic strips. However, these systems does not provide extended user authentication . In other words, being in possession of the card itself and sometimes knowing the pin number is how a user is authenticated [21]. Security is essential for mobile payment systems. Many security standards such as PCI-DSS (Payment Card Industry Data Security Standard) [22], which was first released in 2004, is used to maintain the CIA triad. The people or merchants who use payment cards follow PCI-DSS standards but security violations can still occur [23]. When security violations occur, personal information, payment card information such. With digital wallets, there are several ways to authenticate users such as location [24] and biometrics [25]. Therefore, a new architecture that is tailored for mobile payment and utilizes smartphones' authentication capabilities is needed. There has been several research that proposes mobile payment systems that utilizes a decentralized architecture such has blockchain [26] [27] [28]. However, such decentralized systems does not work well with current banking systems [29]. SWAPEROO [30] proposes a digital wallet architecture that can support more protocols and can be interfaced to several applications. However, it did not discuss adding more authentication factors.

In this paper, a new architecture is introduced that the leverage user authentication capabilities in smartphones to increase the security proximity-based mobile payment systems.

## III. PROPOSED SYSTEM

This section discussed the proposed architecture. As shown in Fig. 3, in the proposed architecture system. The digital wallet App generates a payment token that has several pieces of information that can be used for authentication. This token is forwarded from the POS to the proposed system to be authenticated. In the proposed mobile payment system architecture, multiple nodes of authentication are deployed, where each point receives the payment token and validates its the portion of the token that it is specialized in. Then, it sends the validation outcome to the card network, which makes the final design and send it to the POS.

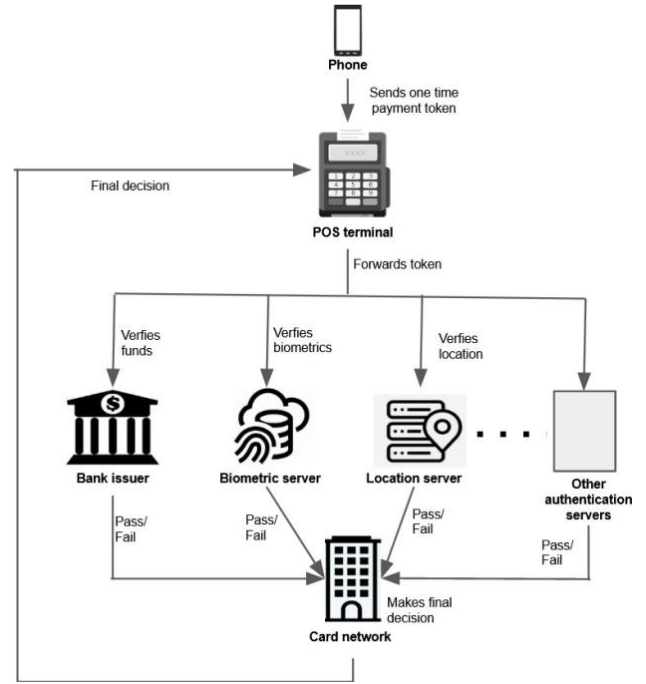

Fig. 3. System Architecture

To explain the transaction process in more detail. The operations are explained in the steps below take place:

1) When the digital wallet is set up it receives authentication tokens from the participating authentication nodes. These tokens will be used in generating payment tokens.
2) The digital wallet generates a token that has the time, biometrics, and location information and possibly other verification a follows:

$$OTTC = LT||BT||FT...||Othertokens \quad (2)$$

Where $LT$ is Location authentication token, $BT$ is biometrics authentication tokens, $FT$ is funds authentication token. $|$ denotes concatenation. Details on the construct of these tokens is discussed below in separate subsections. This token is forwarded to all of the corresponding authentication node.
3) The Digital wallet App sends the token generated in equation 2 to the POS via NFC.
4) This token is forwarded to the participating authentication nodes.
5) Each node verifies there corresponding portion of the token. As will be discussed below, each authentication node will be able to verify its corresponding node as information inside token is hidden using the HMAC function.
6) Each authentication nodes forwards authentication results to the card network.
7) The card network makes the final decision of approving or declining the transaction and sends this decsion to the POS.

More details about how each authentication nodes verify its portion of the OTTC is discussed below.

### A. Fund verification

Fund verification transactions is a process that ensures that there are enough funds available in the account. Usually, fund verification is done by the card issuer. The FT received in yhe payment token is as follows:

$$FT = Amt||t||enc((Amt||t), PK) \quad (3)$$

Where $Amt$ is the transaction amount and $PK$ is the digital wallet private key. The fund verification node. Does the following checks before approving the transaction:

1) Verifies the digital signature of the FT.
2) Verifies the transaction time is within the allowed window
3) Verifies that there are enough funds to cover this transaction

If the checks above succeeds, the authentication passes, else the authentication fails. The authentication result is sent to the card network.

## B. Biometrics verification

Biometrics on smartphones refers to the use of unique physical characteristics, such as fingerprints, facial features, or iris patterns, to authenticate a user's identity. biometrics authentication methods are becoming increasingly popular on smartphones as they provide a convenient and secure way for users to verify their identity. Biometrics in some smartphone today are stored locally [31]. In the proposed architecture, biometrics or information that represents biometrics are stored on a server. This to verify that biometrics submitted by the user in payment token matches the biometrics stored in the node. Biometrics authentication token is as shown below:

$$BT = HMAC(B, HMAC(t, BPS)) \quad (4)$$

Where $B$ is the submitted biometrics at the time of payment. $BPS$ is biometrics pre-shared token. Which is shared by the biometrics authentication node to the digital wallet App to use in future transactions. To verify the biometrics submitted at the time of payment. The biometrics authentication node regenerates the token. By using biometrics submitted for the user during the sign up process. If the token generated at the node equals the token received, the authentication passes, else the authentication fails. The authentication result is sent to the card network.

## C. Location

There are several ways to calculate the location of smartphones (GPS, Wi-Fi, IP address, BLE, etc). In the proposed architecture, The location node has access to the current and previous locations of the device that are obtained using one or more of the localization methods discussed above. The location token is generated in the digital wallet as follows:

$$LT = HMAC(L, HMAC(t, LPS)) \quad (5)$$

Where $L$ is the submitted location at the time of payment. $LPS$ is location pre-shared token, which is shared by the location authentication node to the digital wallet App to use in future transactions. To verify the location submitted at the time of payment. The location authentication node regenerates the token. By using biometrics registers for the user at the sign up. If the token generated equals the token received, the authentication passes, else the authentication fails. The authentication result is sent to the card network.

## D. Other verification

The proposed architecture allows implementing other nodes. For example, behavioral authentication, smart keys, etc. The proposed architecture allows submitting information to be verified in the payment token, when is is forwarded to a cores spending authentication node.

Finally, the card network receives the authentication results from corresponding authentication nodes and makes the final decision of approving or declining the transaction based on the rules configured in the system.

## IV. DISCUSSION

Secure mobile payments are important because they protect sensitive financial information, such as credit card numbers and bank account information, from being intercepted or stolen by hackers. This helps to prevent identity theft and financial fraud. Additionally, secure mobile payments can provide added convenience for consumers, as they can make purchases without having to carry cash or credit cards with them. The proposed architecture provide a methodology for secure verification of multiple elements of a mobile payment. The proposed architecture shows multiple items to be verified and to ensure the multiple aspects to prevent fraud

1) Fund verification: Fund verification is important because it helps to ensure that transactions are completed smoothly and prevents issues like "bounced" payments or chargebacks, which occur when a transaction is completed but the funds are not available to cover it. Additionally, fund verification can also help to prevent fraud by alerting financial institutions if an account has been compromised or if there are any suspicious transactions.
2) Biometrics verification: The use of biometrics is crucial in linking transactions to the individual conducting them. There are various methods available on smartphones and wearable devices, such as FaceID, fingerprint, and blood pressure, to gather biometrics. During enrollment, the biometrics verification node is provided with the user's biometric information. Although "something you have" authentication is rarely utilized in payment systems, linking biometrics to payments ensures that only the payment method's rightful owner can use it, thereby reducing the risk of fraud and identity theft.
3) Location verification: Location verification for credit card transactions is important because it helps to prevent fraud by confirming that the person using the credit card is physically present at the location of the transaction. This can help to prevent "card-not-present" fraud, which occurs when someone uses a stolen credit card number to make a purchase over the phone or online. By verifying the location of the transaction, merchants and financial institutions can be more confident that the person using the card is the legitimate cardholder. Additionally, location verification can also help to prevent "card skimming", which is when a fraudster attaches a small device to a card reader in order to steal credit card information.
4) Other verification's: This allows implementing verification techniques that can optimize the payment process to fit certain scenario for specific cases.

## V. Conclusion

Mobile payment is a fast-growing alternative to traditional payment methods that uses mobile devices for transactions. The current mobile payment systems in digital wallets, are designed to work with existing payment networks but may still pose security risks. A new payment architecture with improved authentication methods is proposed to address these risks. When evaluating mobile payment systems, authentication, encryption, and fraud detection should be considered for maximum security. The mobile payment market is expected to grow significantly in the next decade, highlighting the importance of ensuring security for customers. Secure mobile payments are crucial for protecting sensitive financial information from theft and fraud. The proposed architecture provides a methodology for secure verification of multiple elements of a mobile payment including fund verification, biometrics verification, location verification, and other verification techniques. By verifying these elements, the proposed architecture helps to ensure the legitimacy of transactions, prevent identity theft, and provide added convenience for consumers. This shows that multiple aspects of secure verification are crucial to prevent fraud in mobile payments and enhance the overall payment process.